\documentclass{article}
\usepackage{graphicx}
\usepackage{amssymb}
\usepackage{multicol}
\usepackage{epsfig}
\newcommand{\be}{\begin{equation}}
\newcommand{\ee}{\end{equation}}
\begin{document}

\noindent
{\bf \huge Evidence of a universal power law characterizing the evolution of metabolic networks}

\vspace{0.2cm} \noindent
\renewcommand{\thefootnote}{\fnsymbol{footnote}}
{\bf Shalini$^1$, Areejit Samal$^1$, 
Varun Giri$^1$, Sandeep Krishna$^2$ \footnote[2]{Present address: Niels Bohr Institute for 
Astronomy, Physics and Geophysics, Blegdamsvej 17, Copenhagen DK-2100, Denmark.},
N. Raghuram$^3$ \& Sanjay Jain$^{1,4,5}$ \footnote[1]{Corresponding author (jain@physics.du.ac.in)}}

\vspace{0.2cm} \noindent
{\small \it $^1$Department of Physics and Astrophysics, University of Delhi,
Delhi 110007, India \\
$^2$National Centre for Biological Sciences, UAS-GKVK Campus, Bangalore 560065, India \\
$^3$School of Biotechnology, GGS Indraprastha University, Delhi
110006, India \\
$^4$Jawaharlal Nehru Centre for Advanced Scientific Research, Bangalore 
560064,
India \\
$^5$Santa Fe Institute, 1399 Hyde Park Road, Santa Fe, NM 87501, USA}

\vspace{0.4cm}
\noindent
{\bf 
Metabolic networks are known to be scale free but the evolutionary origin of this
structural property is not clearly understood. One way of studying the dynamical
process is to compare the metabolic networks of species that have arisen at 
different points in evolution and hence are related to each other to varying
extents. We have compared the reaction sets of each metabolite across and within 
15 groups of species. For a given pair of species and a given metabolite,
the number $\Delta k$ of reactions of the metabolite that appear in the metabolic
network of only one species and not the other is a measure of the distance between 
the two networks. While $\Delta k$ is small within groups of related species and 
large across groups, we find its probability distribution to be 
$\sim (\Delta k)^{-\gamma'}$ where $\gamma'$ is a universal exponent that is 
the same within and across groups. This exponent equals, upto statistical 
uncertainties, the exponent $\gamma$ in the scale free degree distribution 
$\sim k^{-\gamma}$. We argue that this, as well as our finding that $\Delta k$ 
is approximately linearly correlated with the degree $k$ of the metabolite, 
is evidence of a `proportionate change' process in evolution. We also discuss some
molecular mechanisms that might be responsible for such an evolutionary process.}

\vspace{0.4cm}
\begin{multicols}{2}

Metabolic networks are known to have a wide, scale free 
distribution of the degree of connectivity of their metabolites \cite{JTAOB2000,WF2001},
$P(k) \sim k^{-\gamma}$ (for a review see \cite{BO2004}). The exponent 
$\gamma$ has been found to have
a value close to 2.2 across all species of organisms that have been studied
\cite{JTAOB2000}\ even though the sets of metabolites and reactions 
in metabolic networks vary quite substantially across organisms.
This structure suggests that some universal process is responsible for the
evolution of metabolic networks; however
at the present time the nature of this process is not clearly understood. 
A comparative study of the metabolic networks of organisms that 
are at varying distances from each other on the evolutionary ladder can shed
light on this process. Here we report on such a study that reveals
some universal features of this dynamical process. 

For growing networks, such as the internet, a preferential attachment of new nodes to 
higher degree nodes \cite{BA1999} as well as a proportionate change mechanism 
\cite{HA1999}
whereby nodes with higher degree experience proportionately higher changes
in degree has been proposed to account for the scale free structure of the network.
The latter process can lead to robust exponents \cite{KPJ2002}. The metabolic network,
however, is not a growing network like the internet; during the course of evolution the
network has changed without a substantial change in the number of metabolites.
Furthermore, so far no concrete evidence has been presented for a preferential
attachment or proportionate change process during its evolution. Here we also
present evidence for a proportionate change process in metabolic network evolution.

We downloaded a database of metabolic networks of  107 organisms \cite{MZ2003}.
This contains organisms from all three kingdoms: eukaryotes, prokaryotes and archaea,
arranged in 15 groups including animals, plants, fungi, proteobacteria, firmicutes, and 
others. Organisms in different groups are evolutionarily distant, while those in the
same group are relatively closeby. We selected one species from each group (typically the 
one having the largest number of metabolites) and compared the metabolic networks 
of all 15 species pairwise (105 pairs of distant species). We also compared specific 
pairs of nearby species (within the same group). 

The metabolic network of a given species of organisms is the set of catalysed chemical 
reactions that can take place in the organism through which it converts
`food molecules' into certain other types of molecules needed by the organism.
The above database contains a list of 5275 metabolic reactions, each reaction characterized
by its chemical equation and whether it is reversible or not. In particular, for every reaction
the list of metabolites participating in the reaction (i.e., the set of 
molecules that are reactants and products of the reaction) is available.
For each reaction and species in the database, information is provided as to whether the 
reaction is present in the metabolic network of that species. The metabolic network
of a given species typically contains several hundred reactions and participating metabolites.

For a given pair of species, say $A$ and $B$, consider the union $M$ of the set of 
metabolites present in each metabolic network. For every metabolite in $M$, we 
now define $\Delta k$, a measure of the distance between the two networks.
Consider a metabolite in $M$,
and let $R_A$ ($R_B$) be the set of reactions in the metabolic network of $A$ ($B$)
in which this metabolite participates. The number $k_A$ ($k_B$) of 
reactions in $R_A$ ($R_B$) is the degree of the metabolite in the species $A$ ($B$). 
Here we consider only the undirected degree of a metabolite, i.e., we do not distinguish
whether the metabolite participates as a reactant or a product. 
Reversible reactions (forward and reverse pair) in which a metabolite participates 
are treated as a single reaction for calculating its degree. If 
a metabolite occurs in only one of the species (say, $A$) and not the other ($B$),
then $R_B$ is the null set and $k_B=0$. 

Consider the reactions in $R_A \cap R_B$. 
The reactions in $R_A \cap R_B$ represent the
links of the metabolite that are common to both species, and hence
$k_{AB}$, the size of this set, is a measure of how much the 
reaction set of this metabolite has remained `conserved' in the 
evolution leading to species $A$ and $B$ from their last common ancestor. 
Similarly, set $(R_A \cup R_B) \backslash (R_A \cap R_B)$, that is, the set of 
reactions in $R_A \cup R_B$ that are not in $R_A \cap R_B$, or 
equivalently those reactions of this metabolite that are in one network
but not the other, measures the divergence between the two networks.
The size of the latter set will be referred to as the divergence of the
reaction sets of this metabolite between species $A$ and $B$, and will
be denoted $\Delta k$. Note that $\Delta k$ is different from the magnitude of 
$k_A - k_B$. For example $R_A$ and $R_B$ can be different sets of reactions 
with the same number of reactions in which case $k_A - k_B = 0$ while
$\Delta k \neq 0$. $\Delta k$ is a measure of the difference between the
two networks that takes into account the {\it identity} of reactions and not
just their number.

We computed the degree distribution $P(k)$ for each of the 15 organisms
as well as the `divergence probability distribution' $Q(\Delta k)$ for each of 
the 105 pairs.
By definition, for any pair ($A,B$), $Q(\Delta k) \equiv n(\Delta k)/|M|$, where
$|M|$ is the number of metabolites in $M$ and $n(\Delta k)$ is the number 
of metabolites in $M$ whose divergence of reaction sets between $A$ and $B$
is equal to $\Delta k$. The divergence probability distribution for 
pairs of distant organisms is shown in Fig. 1 and compared with the degree
distribution. The figure shows that $Q(\Delta k) \sim (\Delta k)^{- \gamma'}$ 
with $\gamma' = \gamma$ upto statistical uncertainties in both the 
exponents. That the degree distribution  of two species follows the power law 
$P(k) \sim k^{-\gamma}$ is a statement of the present structure of the two
metabolic networks. This in no way implies that $Q(\Delta k)$ should
also follow a power law with the same exponent. The latter is a 
distinct statement about the dynamical process that leads to the present structure. 
That the $Q(\Delta k)$ distribution has the same form for all 105 pairs of distant 
species considered reflects a universal property of the evolutionary process.

A comparison of distant species reveals features of the evolutionary process
over long time scales. In order to study the process over short time scales we
compared nearby species (that were in the same group). The result of three 
such comparisons is shown in Fig. 2. As expected, for each pair of nearby species the 
absolute divergence is smaller than for distant species. This is evident from
the fact that the $Q(\Delta k)$ curves are well below the $P(k)$ curves in
Fig. 2, in contrast to Fig. 1 where they are much closer, and that larger
values of $\Delta k$ are absent in Fig. 2. However, it can be seen that the 
$Q(\Delta k)$ still follows a power law with the same exponent as before.
This suggests that this feature of the evolutionary process is also valid over
short evolutionary time scales.

We explored the relationship between $\Delta k$ for a metabolite across
a pair of species, and its degree in each of those species. 
In particular, one can ask for the conditional probability $P(\Delta k | k)$
for a metabolite to have a reaction set divergence $\Delta k$ across a
pair of species, given that its average degree in the two species is $k$.
We found a positive and approximately linear correlation between $\Delta k$ and
the degree of a metabolite (Fig. 3). This is evidence of a `proportionate
change' type process \cite{HA1999}\ in the evolution of metabolic networks.

This provides insight into why the exponents $\gamma'$ and $\gamma$ might be 
equal or very close. For, let us assume for the moment a perfect correlation 
between $\Delta k$ 
and $k$, i.e., $P(\Delta k | k) \sim \delta(\Delta k - f(k))$, or, equivalently, that 
$\Delta k = f(k)$ for some fixed one-to-one function $f$, and also that 
$P(k) \sim k^{-\gamma}$. Then the statement $f(k)\sim k^{\alpha}$ is
equivalent to the statement  $Q(\Delta k) \sim 
(\Delta k)^{-\gamma'}$, with $\gamma' = \gamma / \alpha$. 
In particular $\alpha = 1$ implies $\gamma' = \gamma$ and vice
versa \cite{exponents}. 
However this is not a complete 
explanation because, as is evident from Fig. 3, there is stochasticity in the
relation between $\Delta k$ and $k$, and not perfect correlation. 

What kind of molecular process can give rise to this linear correlation 
between $\Delta k$ and degree of a metabolite? 
A reaction is catalyzed by an enzyme to which the reactant
molecules bind at specific sites in a 3-dimensional geometry.
The structure of the enzyme is determined
by the gene or genes that code for it, and genes evolve via several
mechanisms, including random mutations and gene duplication followed
by divergence through independent mutations in both the copies.
A metabolite with high degree binds to several enzymes
that catalyze its reactions. 
If a gene corresponding to one of
these enzymes mutates in a manner that disturbs the binding site of 
this metabolite on the enzyme, the corresponding
reaction could be lost. The more enzymes the metabolite binds to,
the proportionately higher is the probability of losing its reactions through 
random mutations. On the other hand if the gene
duplicates and diverges, that can introduce a new
enzyme to which it binds and hence a new reaction for this
metabolite to participate in. Large degree metabolites have
a larger pool of interacting enzymes whose genes can duplicate,
and hence if genes duplicate randomly, the number of new reactions a
given metabolite participates in
is also expected to be positively correlated with its degree.
Thus the same mechanisms, namely gene mutations and duplication-
divergence, that have been considered as mechanisms for proportionate
change and preferential attachment in protein interaction networks 
\cite{WOB2003,BO2004}, could 
operate for metabolic networks also.

\noindent
{\bf Acknowledgements:} 
We thank S.N. Bose National Centre for Basic Sciences, Kolkata, and Centre for 
High Energy Physics, IISc, Bangalore for infrastructure and hospitality where part of this
work was done. Shalini and A.S. acknowledge a Junior Research Fellowship from 
UGC and CSIR, respectively.

\end{multicols}

\noindent
{\bf Figure captions}

\noindent
{\bf Fig. 1.} The probability distribution of the divergence of reaction sets 
$Q(\Delta k)$ for evolutionarily distant organisms
compared with the degree distribution $P(k)$ of those organisms. The blue curve 
(joining the blue triangles) in each
figure gives the divergence distribution for the corresponding set of species. For that
curve the x-axis of the figure represents $\Delta k$ and the y-axis represents 
$Q(\Delta k)$. The other
curves give the degree distribution for those species. For these curves the x-axis
represents $k$ and the y-axis represents $P(k)$. Both axes are on a logarithmic
scale in all figures. $k$ and $\Delta k$ are binned logarithmically \cite{logbin}.
Figs. 1a-c compare species from the three
kingdoms pairwise, namely the eukaryote {\it Homo sapiens}, the
prokaryote {\it Escherichia coli} and the archaean
{\it Methanosarcina mazei}. The $P(k)$ curves for these organisms
are given by red dots, green squares and brown hexagons, respectively.
(a) Comparison between $P(k)$ for {\it H. sapiens}
(red dots), $P(k)$ for {\it E. coli} (green squares) and $Q(\Delta k)$ across
these two species (blue triangles). (b) A similar comparison of {\it H. sapiens}
and {\it M. mazei}. (c) A similar comparison of {\it E. coli} and 
{\it M. mazei}.
(d) An average taken over 15 species each drawn from a different group
in the database. For a given metabolite, the average $k$ over the 15 species
and the average $\Delta k$ across the 105 pairs of species are computed
and binned logarithmically. The cyan rhombuses represent 
$P(k)$ and blue triangles $Q(\Delta k)$
in Fig. 1d. The blue lines in Figs. 1a-d are consistent with
$Q(\Delta k) \sim (\Delta k)^{-\gamma'}$. The least square fit value of the 
slope $\gamma'$ $\pm$ standard error arising from the scatter of the points
plotted in the figures is (a) $2.16 \pm 0.13 $, (b) $2.21 \pm 0.10$, (c) $2.23 
\pm 0.08$, (d) $2.13 \pm 0.13$. The values of the exponent in $P(k) \sim 
k^{-\gamma}$ are $\gamma = 2.21 \pm 0.09$ (red), $2.19 \pm 0.11$ (green),
$2.18 \pm 0.12$ (brown), and $ 2.07 \pm 0.11$ (cyan). For each of the 15 organisms 
considered,
$\gamma$ ranges from 2.09 to 2.21 with  mean $\pm$ standard deviation =
$2.16 \pm 0.05$, while for each of the 105 pairs $\gamma'$ ranges from
2.09 to 2.37 with a mean of $2.17 \pm 0.04$.

\vspace{0.1in}
\noindent
{\bf Fig. 2.}
$Q(\Delta k)$ and $P(k)$ for evolutionarily closeby species within the
same group. 
Conventions are the same as for Fig. 1a-c, except that the individual
species are different. Though $\Delta k$ values are smaller for closeby species
as compared to distant species, $Q(\Delta k)$ nevertheless seems 
consistent with a power law with the same exponent as $P(k)$. 
Fig. 2a compares two eukaryotes, both yeasts, {\it Saccharomyces cerevisiae}
($\gamma = 2.09 \pm 0.10$, green), and {\it Schizosaccharomyces pombe} 
($\gamma = 2.18 \pm 0.14$, pink),
and $\gamma'$ is found to be $2.28 \pm 0.10$ (blue). Fig. 2b compares 
two prokaryotes, both proteobacteria, {\it E. coli} ($2.19 \pm 0.11$, green) and 
{\it Salmonella typhimurium} ($2.17 \pm 0.11$, pink); $\gamma' = 2.18 \pm 0.11$ (blue).
Fig. 2c compares two archaea, {\it Pyrococcus horikoshi} ($2.25 \pm 0.10$, green) 
and {\it Pyrococcus furiosus} ($2.17 \pm 0.09$, pink); $\gamma' = 2.28 \pm 0.16$ (blue).

\vspace{0.1in}
\noindent
{\bf Fig. 3.}
Fig. 3. Positive and approximately linear correlation between the divergence 
of the reaction
set of a metabolite and the degree of the metabolite. (a) Scatter plot (on a linear scale) of the 
average $\Delta k$ of a metabolite across the 105 pairs of species versus
its average degree across the 15 (distant) species. The lone point on the 
extreme right is a single highly connected metabolite, the hydrogen ion. 
The plot appears approximately linear with some stochasticity.
(b) The same on a logarithmic scale where metabolites are placed in
logarithmic bins according to their average degree, and the average 
$\Delta k$ for a bin is computed by averaging over all 105 pairs of organisms for
a given metabolite and then averaging over all metabolites in the bin.
The slope of the least square fitted straight line $\pm$ the standard error
of the deviation of points in the figure from the fit is $1.08  \pm 0.03$.
(c). $\Delta k$ versus degree of a metabolite for three pairs of distant species.
The three pairs of species chosen are the same as in Figs. 1a-c. For each 
pair of species ($A,B$), the x-axis represents $k_{\rm max} = {\rm max}(k_A,k_B)$
(the larger of the two degrees of the metabolite in the two species). 
The slopes of the three lines are $1.09 \pm 0.03$ (green) for 
{\it H. sapiens} and {\it E. coli}; $1.08 \pm 0.02$ (pink) for 
{\it H. sapiens} and {\it M. mazei}; and $1.03 \pm 0.02$ (cyan) for {\it E. coli} 
and {\it M. mazei}. (d) $\Delta k$ versus $k_{\rm max}$ 
of a metabolite for three pairs of closeby species.
The three pairs of species chosen are the same as in Figs. 2a-c. 
These have a larger scatter than distant species. 
The slopes of best fit lines are $1.03 \pm 0.07$ (green) for 
{\it S. cerevisiae} and {\it S. pombe};
$0.97 \pm 0.12$ (pink) for{\it E. coli} and {\it S. typhimurium};
and $1.07 \pm 0.08$ for {\it P. horikoshi} and {\it P. furiosus}. 

\begin{figure}[htp]
\vspace{-5cm}
\includegraphics{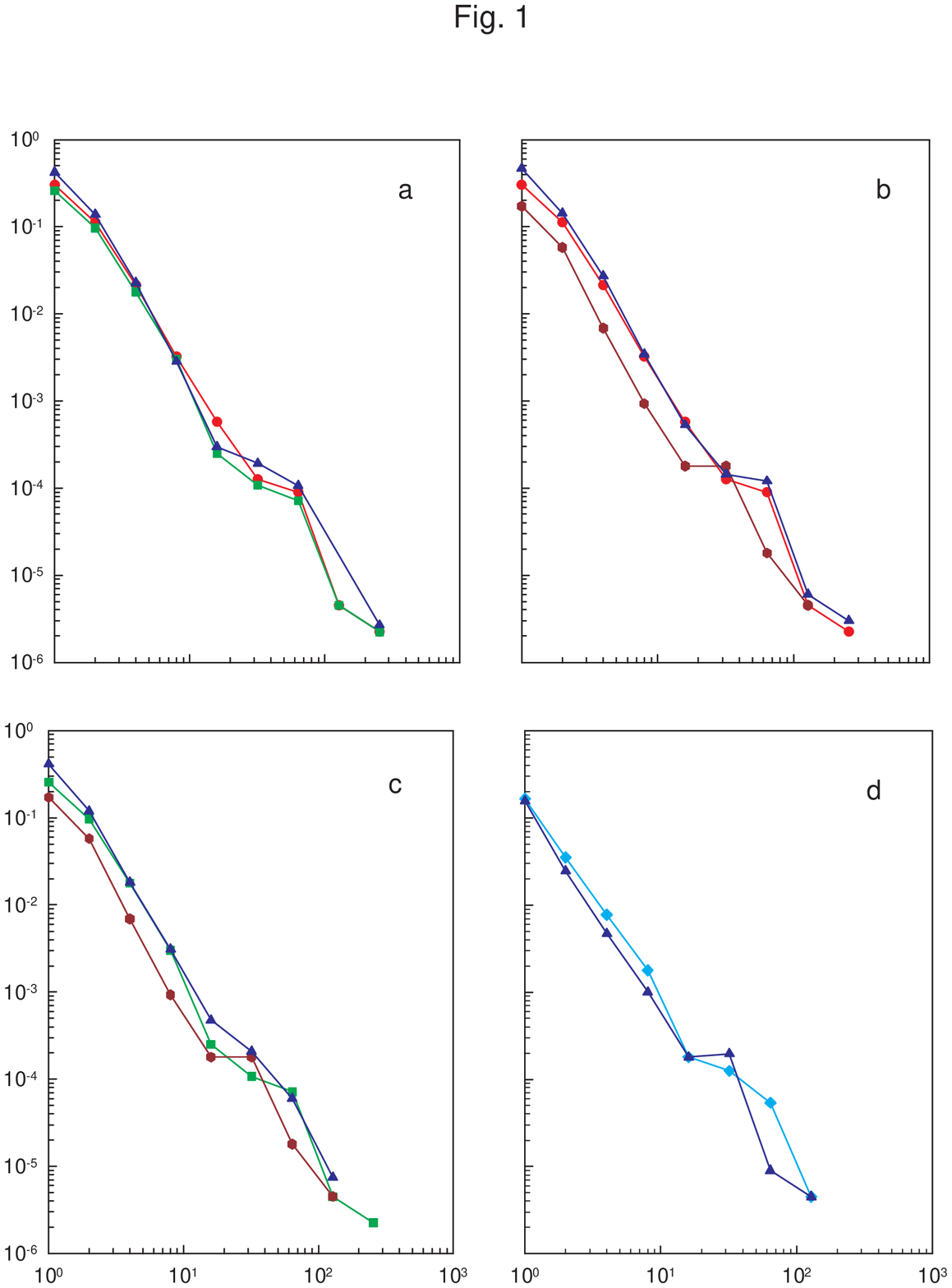}
\end{figure}

\begin{figure}[htp]
\vspace{-4cm}
\includegraphics[angle = 90]{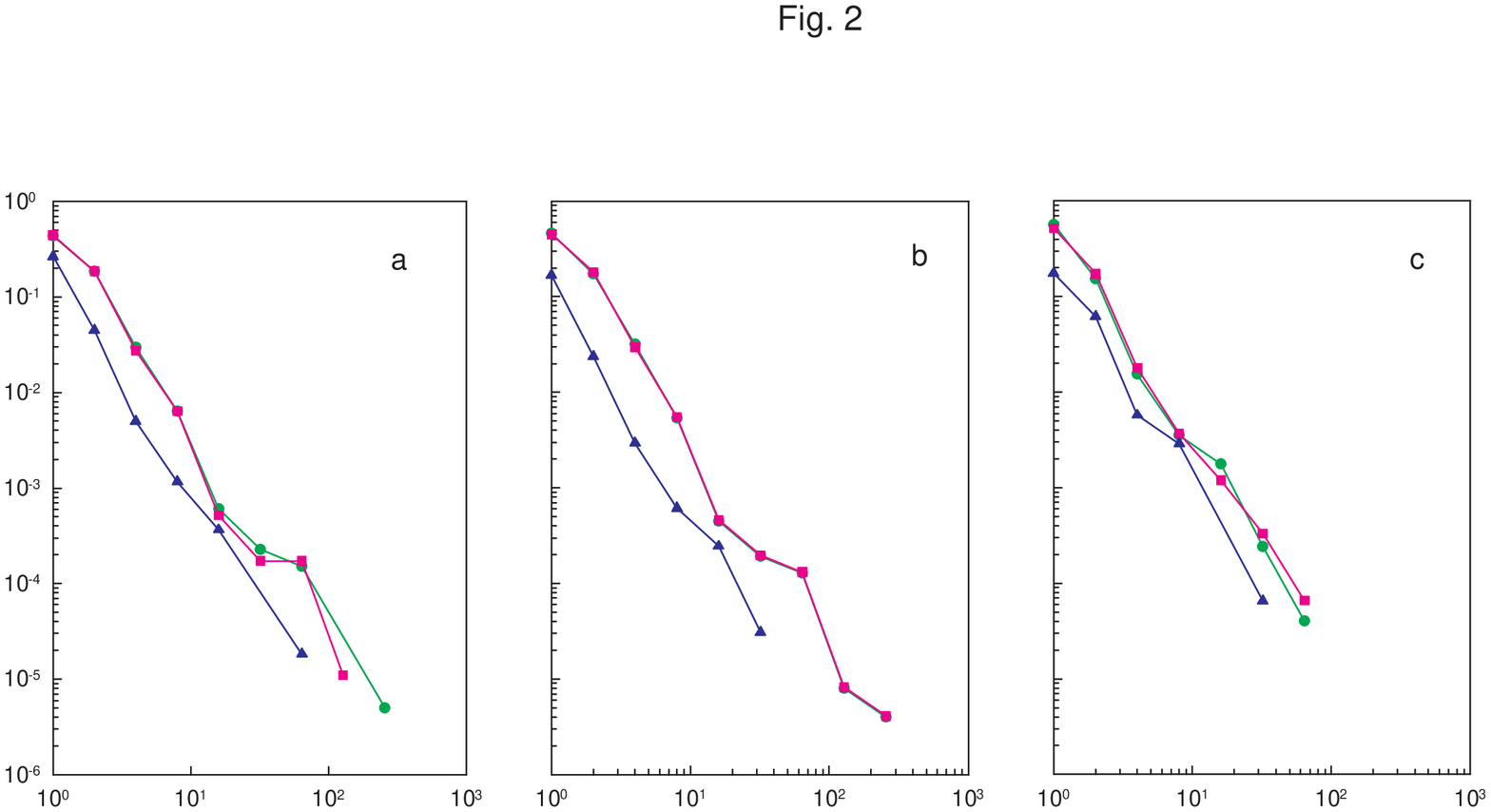}
\end{figure}

\begin{figure}[htp]
\vspace{-5cm}
\includegraphics{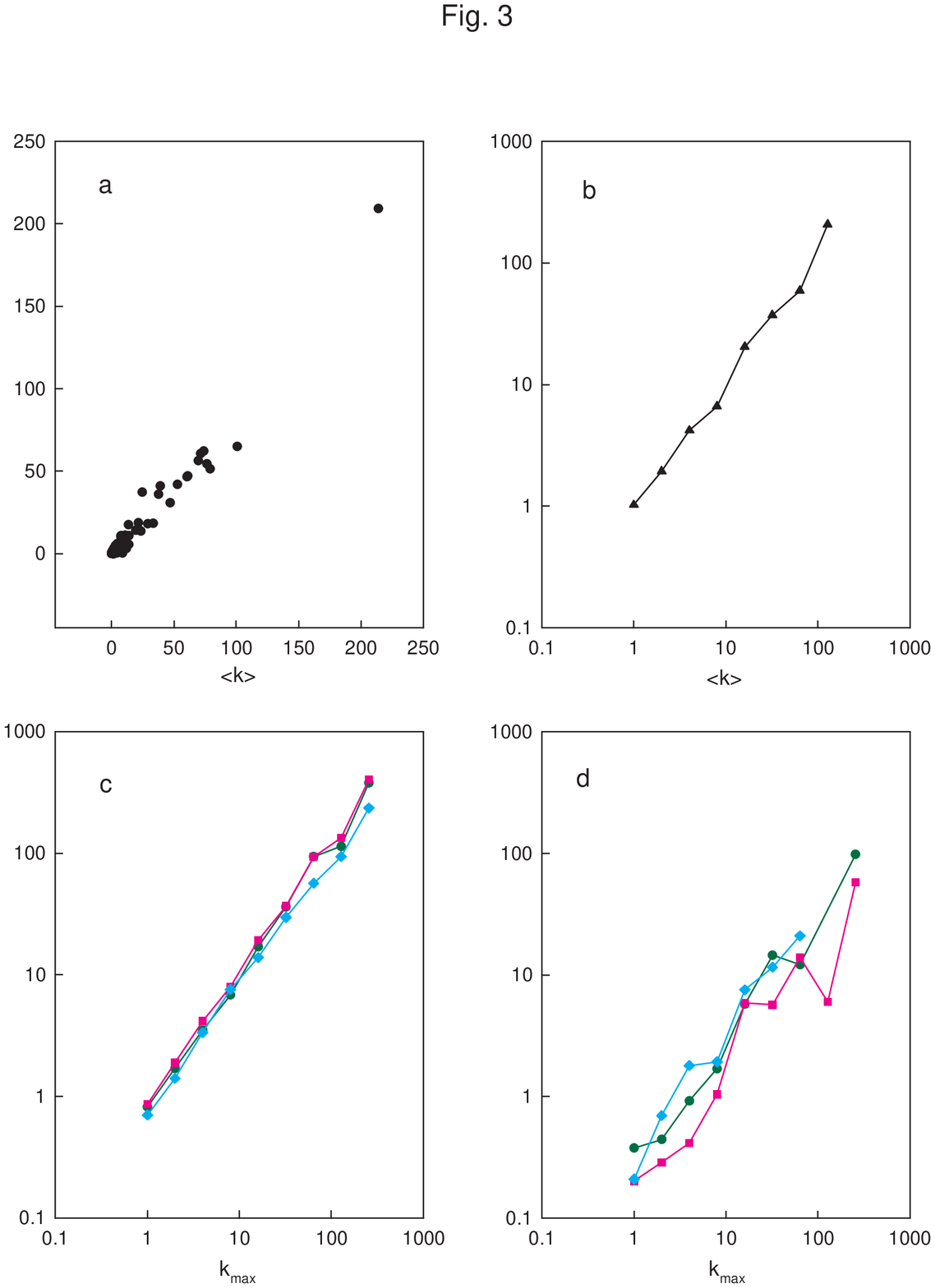}
\end{figure}


{\small
\begin{thebibliography}{99}
\bibitem{JTAOB2000}{H. Jeong, B. Tombor, R. Albert, Z. N. Oltvai, A. -L. 
Barabasi, The large-scale organization of metabolic networks. {\it Nature} 
{\bf 407}, 651-654 (2000).}
\bibitem{WF2001}{ A. Wagner, D. A. Fell, The small world inside large 
metabolic networks. {\it Proc. R. Soc. Lond. B} {\bf 268}, 1803-1810 
(2001).}
\bibitem{BO2004}{A. -L. Barabasi, Z. N. Oltvai, Network Biology: 
Understanding the cell's functional organization. {\it Nature Reviews 
Genetics} {\bf 5}, 101-113 (2004).}
\bibitem{BA1999}{ A. -L.Barabasi, R. Albert, Emergence of scaling in Random 
networks. {\it Science} {\bf 286}, 509-512 (1999).}
\bibitem{HA1999}{B. A. Huberman and L. A. Adamic, Growth dynamics of the
world-wide web, {\it Nature} {\bf 401}, 131 (1999).}
\bibitem{KPJ2002}{B. Kahng, Y. Park and H. Jeong, 
Robustness of the in-degree exponent of
the world-wide web, {\it Phys. Rev. E} {\bf 66}, 046107 (2002).}
\bibitem{MZ2003}{H. Ma and A.-P. Zeng (2003) Reconstruction of metabolic networks 
from genome data and analysis of their global structure for various 
organisms. Bioinformatics. 19 (2): 270-277. }
\bibitem{logbin}{In logarithmic binning of $k$, for example, the 
$r^{\rm th}$ bin ($r = 1,2,\ldots $) is plotted at $(x,y)=(k_r,P_r)$, where
$k_r = 2^{r-1}$, $P_r =  n_r/(N.k_r)$, $n_r$ is the number of 
metabolites with degree in the range $k_r \leq k < 2k_r$, and $N$ is the total 
number of metabolites in the network.}
\bibitem{exponents}{We remark that given the statistical uncertainty in the exponents,
our results are consistent with a value of $\alpha$ slightly different from 1 and 
$\gamma'$ slightly different from $\gamma$.}
\bibitem{WOB2003}{S. Wuchty, Z. N. Oltvai, A. -L. Barabasi, Evolutionary
conservation of motif constituents in the yeast protein interaction network.
{\it Nat. Genet.} {\b 35}, 176-179 (2003).} 


\end{thebibliography}
}
\end{document}